\documentclass[12pt]{article}
\usepackage{amsthm,graphicx,amssymb}

\newcounter{theorem}\theoremstyle{remark}
\newtheorem{remark}[theorem]{Remark}

\newcommand*{\rmd}{\mathrm{d}}
\newcommand*{\e}{\mathbf e}%
\newcommand*{\eqref}[1]{(\ref{#1})}
\newcommand*{\tfrac}[2]{\textstyle{\frac{#1}{#2}}}

\DeclareSymbolFont{iso}{U}{txmia}{m}{it}
\DeclareMathSymbol{\rea}{\mathalpha}{iso}{"92}
\DeclareMathSymbol{\okr}{\mathalpha}{iso}{"93}
\DeclareMathSymbol{\nab}{\mathalpha}{iso}{"6E}
\DeclareMathSymbol{\isom}{\mathalpha}{iso}{"24}

\date{}
\title{Falling into the Schwarzschild black hole. Important details.}
\author{S. Krasnikov\thanks{The Central Astronomical Observatory of RAS,
M-140, Pul\-ko\-vo, St.~Petersburg, Russia. \emph{Email}:
Gennady.Krasnikov@pobox.spbu.ru}}
\begin{document}
\maketitle

\begin{abstract}
The Schwarzschild space is one of the best studied spacetimes and its
exhaustive considerations are easily accessible. Nevertheless, for some
reasons it is still surrounded by a lot of misconceptions, myths, and
``paradoxes". In this pedagogical paper an attempt is made to give a
simple (i. e., without cumbersome calculations) but rigorous
consideration of the relevant questions. I argue that 1) an observer
falling into a Schwarzschild black hole will
\emph{not} see ``the entire history of the universe" 2) he will
\emph{not} cross the
horizon at the speed of light 3) when inside the hole, he will \emph{not}
see the (future)  singularity, and 4) the latter is \emph{not}
``central".
\end{abstract}

\section{Introduction}

The Schwarzschild spacetime (alias maximally extended Schwarzschild
spacetime, alias the Kruskal spacetime) is certainly one of the best
studied solutions of the Einstein equations. A rare  textbook in
relativity does not dwell on that space, which is no surprise taking into
account its importance and (relative) simplicity. So, one might think
that no mysteries are harboured there any longer, a careful reading of
\cite[\S\S 31,32]{MTW} being able to give the answer to almost any
``silly" question. This, however, is not quite so. For a person who has
not yet got used to the basic concepts of general relativity (equivalence
of all coordinate systems, the impossibility of attaching a preferred
extended reference system to an observer, etc.) the Schwarzschild space
is fraught with pitfalls. Such a person encounters various ``paradoxes"
and ``miracles" (exactly as in studying special relativity or quantum
mechanics) and it takes some work to sort them out. Unfortunately, the
areal of those paradoxes and miracles is not restricted to student
internet forums and  popular literature. They have infiltrated the
semi-popular, research, and even pedagogical works. Thus one can find
there the assertion that, just before  crossing a black hole horizon, an
astronaut in a single moment of his proper time will see the whole
infinitely long evolution of the external universe \cite{Regge,Cher}. He
will see how our Sun swells becoming a red giant, how the Earth skimming
over the upper atmosphere of the dying Sun evaporates in its glare, and
how the Sun later transforms into a white dwarf \dots\cite{Cher}.
Elsewhere one reads that the astronaut  will traverse the horizon at the
speed of light
\cite{KisLogMes6} and after crossing the horizon he will see the
``central singularity" \cite{Cher}. The authors of these excerptions are
all scholars of authority, so one can only pity a student reading all
that.

Thus, it seems there is a need for a paper where the most puzzling
properties of the Schwarzschild space would be illuminated in an as clear
(but rigorous) manner as possible. In the following sections I treat |
hopefully just in that manner | a few  most ``controversial" issues,
which are: Will an observer falling into the black hole see the entire
future of our universe? Will he cross the horizon at the speed of light?
Will he see the singularity? Is that singularity ``central"? (The answers
to all four questions are negative). The reader is supposed to be
familiar with only the basics of semi-Riemannian geometry. Units are used
in which $G=c=1$.
\section{The locale}
\subsection{The geometry of the Schwarzschild spacetime}
\begin{figure}[tb]
\includegraphics[width=\textwidth]{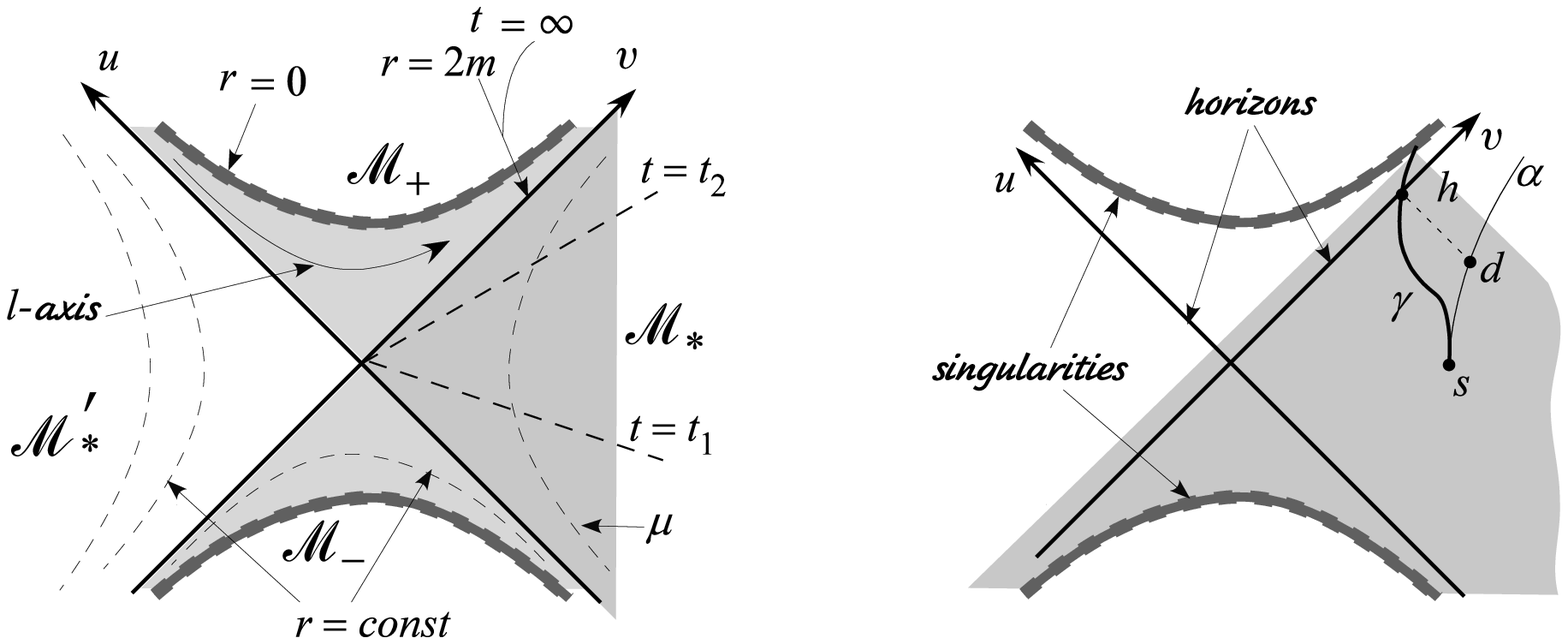}\\
\hspace{0.15\textwidth}(a)
\hfill (b)\hspace*{0.15\textwidth}
\caption{The sections $\phi=const$, $\theta=const$ of the
Schwarzschild spacetime. (a). The dark gray region $\mathcal M_*$
 is asymptotically flat, it is the region
``outside of the Schwarzschild black hole''. By the light gray the
regions $\mathcal M_-$ and $\mathcal M_+$  are shown, which are,
respectively, expanding  and contracting ``universes"; (b). The gray
region is the causal past of $\gamma$, i.~e., the union of the causal
pasts of all its points.\label{fig:Sch} This region includes \emph{all}
events that have ever been observed by $\gamma$. }
\end{figure}

The simplest (i.~e., non-ro\-tat\-ing and uncharged) black hole is
described, as everybody knows, by the spacetime $\mathcal M$:
\begin{equation}\label{eq:metr}
\rmd s^2=
4m^2\left\{-\frac{4}{xe^{x}}\rmd u\rmd v
 +x^2(\rmd\theta^2 + \cos^2\theta\,\rmd\phi)\right\},
\end{equation}
$$
u,v  \in \rea,\quad x>0,
$$
 where $m$ is a positive
parameter (called the mass, see below) and $x=x(u,v)$
 is the function defined (implicitly) by the equation
\begin{equation}\label{eq:def r}
uv=(1-x)e^x.
\end{equation}
The importance of  $\mathcal M$ | it is this spacetime that we shall call
Schwarzschild's | lies, of course, in the fact that it is a spherically
symmetric solution to the vacuum Einstein equations and, moreover, by
Birkhoff's theorem it is the only such solution in the class of
maximal\footnote{ $\mathcal M$ cannot be extended, say, to the region
$x(u,v)<0$ because the scalar $R_{abcd}R^{abcd}$ diverges at $x\to 0$.}
globally hyperbolic spacetimes. It is often convenient to choose $x$ (or
$r$, which is almost the same) as a new coordinate. That cannot be done
in the entire $\mathcal M$ (as follows, for example, from the fact that
$\nabla x(0,0)=0$) and we shall restrict ourselves to the region
\[
\mathcal M_*\colon\qquad u<0,\quad v>0.
\]
(in Fig.~\ref{fig:Sch}a it is shown by dark gray).
There the transition to the coordinates
\begin{equation} \label{ScKoordI}
r\equiv 2mx,\qquad t\equiv 2m\ln(-v/u)
\end{equation}
brings the metric~(\ref{eq:metr}) to a more customary form:
\begin{equation}
\label{eq:UsSch}
 \rmd s^2=-(1-\tfrac{2m}{r})\rmd t^2 + (1-\tfrac{2m}{r})^{-1}\rmd
r^2
 +r^2(\rmd\theta^2 + \cos^2\theta\,\rmd\phi)
\end{equation}
$$
t\in \rea,\quad r>2m.
$$
To relate it to the everyday  consider the region $\mathcal
M_{r_0}\subset \mathcal M_*$ defined by the inequality $r>r_0>2m$.  The
region is spherically symmetric and asymptotically flat, and the metric
there solves the source-free Einstein equations. So, $\mathcal M_{r_0}$
(and | again by Birkhoff's theorem | only $\mathcal M_{r_0}$) describes
the universe outside a ball of radius $r_0$ and mass $m$. (The equation
for the $r$-coordinate of a radial geodesic parametrized by the proper
time $\tau$ is
\footnote{It is easily found by varying the ``geodesic Lagrangian"
\cite{MTW} $L= -(1-
\frac{2m}{r})\dot t^2+ (1-
\frac{2m}{r})^{-1}\dot r^2$ with respect to $r$.}
\begin{equation}\label{eq:radg}
  \ddot r = -\frac m{r^2},
\end{equation}
where the dot stands for the derivative by $\tau$. The comparison of this
equation with the Newtonian one justifies our interpretation of $m$ as
the mass).

The surfaces $u=0$ and $v=0$ (alias $x=1 $, alias $r=2m$) bounding
$\mathcal M_*$ are called \emph{horizons}. It should be emphasized that
the points of horizons  have no ``magic'' properties; each of them has a
small neighbourhood with exactly the same (in a qualitative sense)
properties as a neighbourhood of any other point of any spacetime: the
tidal forces here are finite,  massive bodies move on timelike curves,
the world lines of photons are null geodesics, etc. Now, what is there
\emph{beyond}  the horizon? One might naively expect that since the horizon
is a sphere (at each moment of time; we are discussing the section of the
spacetime by some simultaneity surface $\mathcal S$), then what it bounds
is a ball. Or rather a punctured ball, with a singularity at the center.
That would perfectly fit the idea that the Schwarzschild solution
``describes the field of a point mass (located at the center, the
singular point of the metric)" \cite{LL_II}. The said idea goes back to
classics of the pre-Kruskal epoch \cite{konc} and is amazingly widespread
even today. It should be stressed therefore that the just drawn picture
though not \emph{wrong} ($\mathcal S$ can be chosen so as to justify it)
is, nevertheless, grossly misleading. We shall see, in particular, that
the term ``central" is applicable to Schwarzschild's singularity no more
than, say, to Friedmann's.

To perceive the real  geometry of the region
\[
\mathcal M_+\colon\qquad u>0,\quad 0<r<2m,
\]
(shadowed in light gray in Fig.~\ref{fig:Sch}a) it is instructive to
introduce there the coordinates
\begin{equation} \label{ScKoordII}
\eta\equiv 2mx,\qquad l\equiv 2m\ln(v/u).
\end{equation}
The metric then takes the form
$$
\rmd s^2=-(\tfrac{2m}{\eta}-1)^{-1}\rmd\eta^2 +
(\tfrac{2m}{\eta}-1)\rmd l^2
 +\eta^2(\rmd\theta^2 + \cos^2\theta\,\rmd\phi)
$$
$$
l\in \rea,\quad \eta\in (0,2m).
$$

\begin{remark}
The transformation \eqref{ScKoordI} is singular at $u=0$ and $v=0$.
Therefore it \emph{cannot} be extended to $\mathcal M_+$. In other words,
$(t,r)$ and $(\eta, l)$ are
\emph{different} coordinates. Unfortunately, this fact is overlooked
sometimes, which leads to much confusion and the talk about  ``space and
time swapping their roles'' inside the black hole.
\end{remark}
Thus, an observer after crossing the horizon finds himself in a
``universe'' with not quite usual properties\footnote{This homogeneous,
anisotropic universe is a special case of the Kantowski-Sachs
 spacetime \cite{KS}.}. The ``space'' of that universe (i.~e., the surface
$\mathcal S$ given in this case by the equation $\eta={\rm const} $) is a
homogeneous cylinder $\rea^1\times\okr^2$. It is spherically symmetric,
but not isotropic, the distinguished direction being that along the
$l$-axis. At the same time, even though the surfaces $\eta={\rm const}$,
$l={\rm const}$ are spheres, one should not call the $l$-coordinate
``radial", because the space is invariant w.~r.~t. translations in that
direction. Note that the space has \emph{neither} a singularity,
\emph{nor} a centre.

With time the geometry of $\mathcal M_+$ changes. This fact is not
surprising --- the Schwarz\-schild space as a whole is \emph{non-static},
even though it has a static  [as is seen from
\eqref{eq:UsSch}] region\footnote{The isometries
$\isom^*_A\colon$
$
 t\mapsto t + A,
$
that act on $\mathcal M_*$ \emph{can} be extended to the isometries
$\isom_A\colon$ $u\mapsto e^{-\frac{A}{4m}}u$, $v\mapsto
e^{\frac{A}{4m}}v$ acting on the entire $\mathcal M$, but in  $\mathcal
M_+$  the orbits of the group $\isom_A$ are spacelike.}
 $\mathcal M_*$. The radius of the cylinders $\mathcal S$ falls and
 it is its
vanishing at $\eta=0$ that is referred to as the Schwarzschild
singularity\footnote{The case in point is the ``upper singularity" in
Fig.~\ref{fig:Sch}; the other one is, of course, in the past.}.
Evidently,
 for any observer in $\mathcal M_+$ the singularity is \emph{in the
 future} and, in
 particular, \emph{nobody}
 (on whichever side of the horizon) can ever observe it.

\begin{remark} The surfaces of simultaneity could be chosen
differently, of course. For example, they could be defined by the
equation $u+v={\rm const}$ instead of $\eta={\rm const}$. In such a case
$\mathcal M$ would appear as an evolving wormhole \cite{MTW}. The throat
of the wormhole lies in $\mathcal M_\pm$ and connects two asymptotically
flat isometric ``universes" --- ours and $\mathcal M_*'$. As can be seen
from Fig.~\ref{fig:Sch}, the two universes are causally disconnected, but
a traveller from one of them is allowed to see some events in the other
(though not before traversing the horizon).
\end{remark}

Often it is $\mathcal M_*$ that is called  Schwarzschild's space and $t$,
$r$ --- Schwarzschild's coordinates, while $\mathcal M$ and $u,v$ are
called Kruskal--Szekeres'. The   coordinates  $u,v$  cover the entire
manifold. And, in studying the \emph{radial} motion, when only the
sections $\phi=const$, $\theta=const$ matter,  their additional advantage
is that the metric of those sections takes the form
\[
\rmd s^2=
-F(r)\rmd u\rmd v,\qquad
 F= 16m^2x^{-1}e^{-x},
\]
which simplifies significantly the analysis of their causal structure. A
curve in the $(u,v)$-plane is causal (i.~e., can be the world line of a
particle) if and only if in all its points the angle between its tangent
and the vertical  (i.~e., the line $u-v=const$) is $\leq 45^\circ$. Thus
the set of all points from which signals can come to a point $p$ | this
set is called the \emph{causal past} of $p$ | is the down-directed angle
with the vertex in $p$ and the sides parallel to the $u$- and $v$-axes.
And the  \emph{causal future} of $p$, i.~e., the set of all points at
which $p$ can be seen, is the angle vertical to that.

We shall consider only the  $(u,v)$-plane taking into account that the
non-radial motion complicates the analysis without adding anything
qualitatively new. So, by a ``signal" or ``motion", etc., from now on we
understand a ``radially propagating signal" or ``radial motion", etc.
\subsection{Schwarzschild  and free-falling observers}

To analyze the fall  into the black hole let us consider two observers
separating into a point $s$, see Fig.~\ref{fig:Sch}b. One of them, let us
label him $\alpha$, after the parting moves with constant $r$, $\phi$ and
$\theta$. Such observers | we shall call them Schwarzschild | are at rest
in the Schwarzschild coordinates, in which the metric does not depend on
time. It would be quite untrue, however, to regard Schwarzschild
observers as ``immobile'' or, at least, inertial. Their world lines are
\emph{not} geodesics; so the observers experience an acceleration $a$, the fact
well known (empirically) to the reader as all of us are, to high
accuracy, Schwarzschild observers with $r=R_{\oplus} $ in the metric with
$m=M_{\oplus}$ and each of us moves | in the instantaneously comoving
system | with the acceleration $a\approx 9.8\,$m/s$^2$.

The second observer, $\gamma $, falls freely, i.~e., his world line is a
radial geodesic $\gamma(\tau)$ with $\dot r(0)\leq 0$. The most important
fact about $\gamma $ is that at some moment $\tau_h<\infty$ of its proper
time it
\emph{unavoidably} meets the horizon.
\begin{proof}
As follows from \eqref{eq:radg} the function $x(\tau)$ is convex. At the
same time
\[
x(0)>1,\qquad \dot x(0)\leq 0.
\]
Hence, there \emph{is} $\tau_h>0$ such that $x(\tau_h)=1$. So, we only
have to prove that $\tau$ takes
\emph{all} values in $[0,\tau_h]$.
In other words, the observer $\gamma$ \emph{must} reach the horizon, if
he lives long enough, and our task is to prove that he does not cease to
exist before his clock shows $\tau_h$. Note that this follows neither
from \eqref{eq:radg}, nor from any general considerations: one could
imagine, for example, that $\gamma$ approaches the horizon like $\mu$ in
Fig.~\ref{fig:Sch}a and leaves $\mathcal M_*$ as $\tau\to a\in
(0,\tau_h]$. To exclude such a possibility notice that as long as $\gamma
$ stays in $\mathcal M_*$ the coordinate $v$ on it obeys the following
assessment
\begin{equation}\label{eq:ocen}
\ddot v=-(\ln F),_v{\dot
v}^2=(1-x^{-2})v^{-1}{\dot v}^2 <v^{-1}{\dot v}^2,
\end{equation}
where the first equality is the $v$-component of the geodesic equation
and the second follows from the simple chain
\[
(\ln F),_v=(\ln F)'(vu),_v/(vu)'=v^{-1}\ln'F /\ln'|vu|=v^{-1}(1-x^2)/x^2,
\]
in which we have made  use of \eqref{eq:def r}. Both $v$ and $\dot v$ are
positive in $\mathcal M_*$, and from
\eqref{eq:ocen} it follows immediately that
\[
v(\tau) \leq c_1e^{c_2e^{\tau/c_2}},
\]
where $c_{1,2}$ are some constants. Consequently, until $\gamma $ leaves
 $\mathcal M_*$, $v(\tau)$ is bounded on any interval.
\end{proof}
Once $\gamma $ enters $\mathcal M_+$ it cannot cross the horizon back and
inevitably terminates at the singularity ($r\to 0$ as
$\tau\to\tau_0<\infty$). Note that the same is true for \emph{any} causal
curve | geodesic or not | just because it has to stay within the right
angle with the vertex in its (arbitrary) point and the sides parallel to
the $u$- and $v$-axes.

All the abovesaid looks quite elementary. However, for the reasons
discussed in the Introduction we should discuss in more detail two
aspects of $\gamma$'s history.

\section{The velocity at the horizon}

It is common knowledge that an object \emph{similar to} the black hole
exists in   Newtonian physics too. If a ball has a sufficiently large
mass and small radius, the escape velocity $V_e$ may equal the speed of
light. But a body falling on such a ball  | with the zero initial speed |
from infinity would land just with $V_e$. Perhaps, it is such reasoning
that gave rise to a popular belief that a body crosses the horizon at the
speed of light. Is it true?

At first glance | yes. Indeed, consider a family of observers
 $\mathcal N $: each member $\nab_\tau$ meets $\gamma$
 in the corresponding point | in $\gamma(\tau) $ | and measures
 $\gamma$'s velocity in his, member's, proper reference system.
By the proper reference system we here understand a perfectly local and
well-defined entity | an orthonormal tetrad in $\gamma(\tau) $
\[
\{\e_{(i)}(\tau)\},\qquad  i=0,\dots,3,
\]
with the vector $\e_{(0)}$ tangent to the world line of $\nab_\tau$
(thus, instead of a family of observers we could speak about a tetrad
field along $\gamma $). Denote now by $\mathbf v(\tau)$ the 3-velocity of
$\gamma $ as measured by $\nab_\tau$, i.~e.,  found in the basis
$\{\e_{(i)}(\tau)\}$. If $\mathcal N $ is chosen (at $\tau<\tau_h$, of
course) to be the set of   Schwarzschild observers, then for a radial
$\gamma $ it can be shown, see \cite[(102.7)]{LL_II}, that
\[
|\mathbf v|=\sqrt{1-\xi (x-1)/x},
\]
where $\xi$ is a positive constant which depends on the choice of $\gamma
$. Thus
    \begin{equation}\label{eq:lim} |\mathbf
v|\rightarrow 1 \quad \mathrm{as}\quad \tau\to\tau_h-0.
\end{equation}
It is this fact that is interpreted sometimes as attainment of the speed
of light by a  falling body and thereby as self-inconsistency of general
relativity, see, e.~g., \cite{KisLogMes6}.

The falseness in that interpretation is that $|\mathbf v|$ is assumed to
be continuous in $\tau$. In fact, however, the properties of $|\mathbf
v|$ depend heavily on the choice of $\mathcal N $ (in this sense $\mathbf
v(\tau)$ characterizes $\mathcal N $ rather than $\gamma $). And in the
case under consideration, when $\mathcal N $ can\emph{not} be
complemented in a continuous way by an observer  meeting $\gamma $ in $h$
(such an observer would have to move with the speed of light), the vector
\[
\e_{(0)}=(v\partial_v-u\partial_u)/|v\partial_v-u\partial_u|
=
(v\partial_v-u\partial_u)/\sqrt{32m^2(1-1/x)}
\]
has obviously no limit as $\tau\to\tau_h-0$. So, $\mathbf v(\tau)$ could
have been continuous in $\tau_h$ only by a miracle. In other words,
\eqref{eq:lim} \emph{does not imply} $|\mathbf v(\tau_h)|=1$.

Actually the vector in $h$ tangent to $\gamma $ is \emph{timelike}. This
has nothing to do with relativity, or even with the  metric under
discussion, but follows from a fundamental geometric fact: a geodesic
timelike in a point ($s$ in this case) is timelike in \emph{all} points.
Thus in any orthonormal basis (i.~e., in a proper reference system of any
observer located in $h$) $\gamma$  crosses the horizon moving slower than
light.

\section{What will the falling observer see?}
Another widely met statement is ``From the point of view [or `in the
reference system', or `as measured by the clock'] of a remote observer it
takes infinite time for a body to reach the horizon". In this section I
argue that contrary to the first impression it is \emph{possible} to give
a meaning to that statement and even in three different ways (and,
indeed, in the literature all three meanings can be met). The statement
deserves a detailed analysis, because one of the three interpretations is
simply
\emph{wrong}.

The problem,  in essence, is that  an observer's clock measures the
observer's \emph{proper} time $\tau$, i.~e., for an observer with the
world line $\alpha=x^i(\xi)$ it measures the quantity
\[\tau(\xi)=\int_0^\xi\sqrt{g_{nk}\dot{x}^n\dot{x}^k}\,\rmd\xi'
\]
(the dot here is a derivative with respect to $\xi'$) and no reasonable
way is seen to make the clock measure a time interval between events
lying
\emph{off} $\alpha$ (in our case those are the events  $s$ and $h$).
Normally this causes no problems, because we can pick \emph{any}
coordinate system and measure the time by using it. In doing so we need
not bother to interpret the thus defined time as ``true", or ``measured
by the clock of this or that  observer". Hence, the first way to
interpret the above-mentioned statement is to reformulate it: ``It takes
infinite Schwarzschild time for a body to reach the horizon", or, more
strictly (since the Schwarzschild coordinates do not cover $h$):
``Between $s$ and $h$ there are events on $\gamma$ with arbitrarily large
$t$". The latter statement is trivially true, see Fig.~\ref{fig:Sch}.

\begin{figure}[tb]
\includegraphics[width=\textwidth]{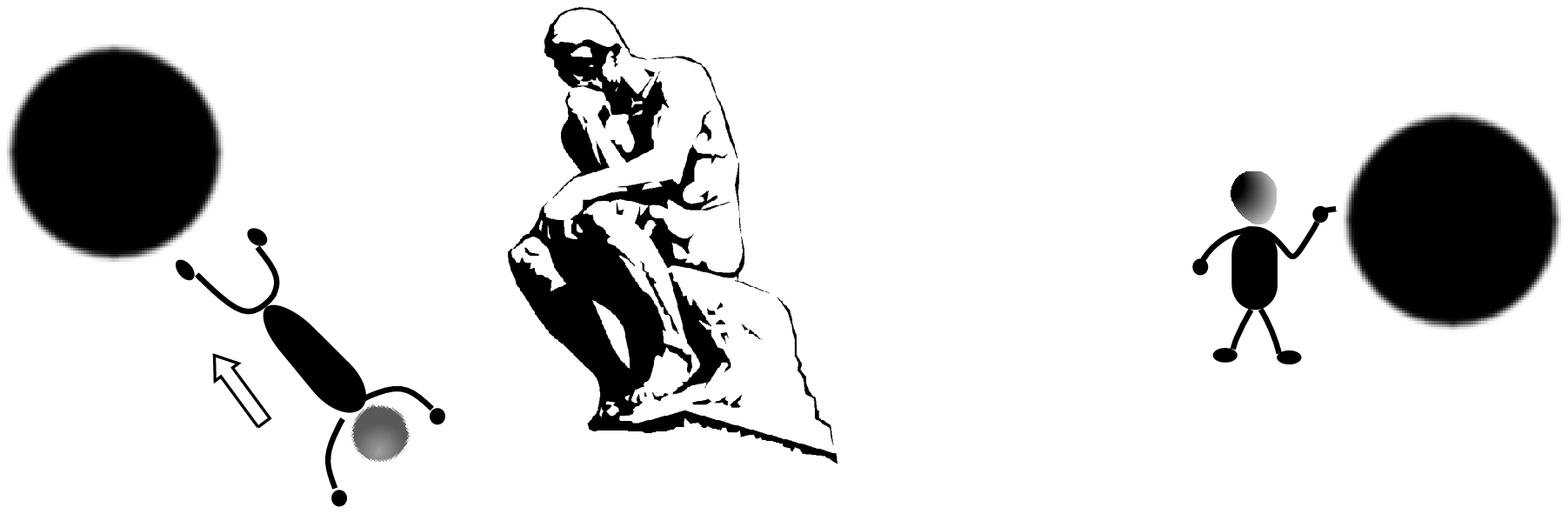}\\[\medskipamount]
\hspace*{0.3\textwidth}(a)
\hfill (b)\hspace*{0.15\textwidth}
\caption{(a). The waist of a free falling victim must cross the
horizon at some moment. But his head at that moment still makes a regular
observer (it moves slower than light w.~r.~t.\ a Schwarzschild observer).
So, does it mean that in ``the head's reference system" the feet have not
yet reached the horizon? (b). ``From the point of view" of a
Schwarzschild observer with the coordinate $r_0$ his distance to the
horizon is $\Delta=\int_{2m}^{r_0}(1-\frac{2m}{r})^{-1/2}\rmd r$.
Evidently $\Delta\rightarrow 0$ as $r_0\to 2m$ and hence at some $r_0$ he
will be only, say, 10$\,$cm far from it. What can prevent the observer
from simply stretching a hand and touching the horizon?
\label{fig:touch}}
\end{figure}

\begin{remark} Replacement of the words ``in the Schwarzschild
coordinates" with ``from the point of view of a remote observer" is quite
common. The point is that light signals sent by a Schwarzschild observer
at   regular intervals $\Delta\tau$ of his proper time are received by
another Schwarzschild observer also at regular intervals\footnote{Because
$\mathcal M_*$  is static.} $\Delta\tau'$ (whatever are the corresponding
intervals $\Delta t$ are $\Delta t'$ of the coordinate time). Which
enables one to ``synchronize the clocks" throughout the entire $\mathcal
M_*$, i.~e., to introduce the time coordinate by requiring that $\Delta
t=\Delta t'$ (and that is how the Schwarzschild time can, indeed, be
defined). The similarity of this procedure to that used for building a
reference frame in special relativity (a purely illusive similarity, of
course, since $\Delta\tau\neq\Delta\tau'$ even though the observers are
``at rest" w.~r.~t.\ each other) can mislead one into the idea that the
Schwarzschild coordinates are ``more physical" than the others and, in
particular, an interval of $t$ is exactly ``the time by the clock of a
remote observer". It is this deeply non-relativistic idea that makes |
 actually simple | properties of $\mathcal M$ look paradoxical. One such
paradox has been already considered, another is considered below, and two
more are presented in Fig.~\ref{fig:touch}.
\end{remark}

The fact by itself that two events are separated by an infinite
coordinate interval is vapid: it is true for \emph{any} pair of causally
related events if the time coordinate is chosen appropriately. A more
meaningful --- since geometric --- statement can be made if we turn from
a relation between two events ($h$ and $s$) to a relation between an
event and an observer ($h$ and $\alpha $, respectively), or two observers
($\alpha $ and $\gamma$). Indeed, notice that the
\emph{whole}
 $\alpha$ lies in the causal future of the \emph{segment} of $\gamma$
 bounded by $s$ and $h$. Physically this means that all his | infinite | life
 $\alpha$ will be able to receive signals sent by his falling comrade
 \emph{before} the latter reached the horizon\footnote{The signals though
 must be sent more and more frequently and with more and more blue
 photons.}. $\alpha$ may interpret this fact in two ways depending on
 which spacelike surfaces he chooses as the surfaces of simultaneity
 (recall that in general relativity there is \emph{no} preferred choice):
\begin{enumerate}
  \item If the surfaces are more or less horizontal in Fig.~\ref{fig:Sch}
  (for instance,   events are regarded simultaneous if they have the
  same value of $u+v$), then $\alpha$ will find nothing unusual in
  receiving the messages from $\gamma$. Exactly as we speak of the
  light of a distant star  coming to us for years after the star died,
  $\alpha$ could say that due to a huge | and growing | delay he keeps
  receiving
  signals from $\gamma$ centuries after the latter \emph{actually}
  traversed the horizon.
  \item One can choose, however, the simultaneity surfaces to be more and
  more tilted (cf.~the surfaces $t=const$ in Fig.~\ref{fig:Sch}a). This |
  rather exotic | choice would mean that $\alpha$ considers the
  information received with every signal as more and more fresh. And he
  would be quite consistent claiming that the fall  is infinitely long.
\end{enumerate}

There is, however, another | opposite, in a sense | approach to what
should be called infinitely long by a remote observer's clock. Imagine a
point $p$ which contains  the \emph{entire} $\alpha $ in its causal past.
An observer $\omega $, if his world line passes through $p$, would see
the entire (infinite) history of $\alpha $: he will see $\alpha $ aging,
his sun swelling and reddening, its protons decaying, etc. In his turn,
$\alpha $ would be able at any moment to send a message to $\omega $
(though maybe not to receive a response). All in all it would be quite
legitimate to say that $\omega $ reaches $p$ in infinite time by
$\alpha$'s clock.

In this just formulated sense the statement that $\gamma$'s falling time
is infinite is \emph{wrong}. Indeed, as is seen from Fig.~\ref{fig:Sch}b,
$\alpha $ | like any other Schwarzschild observer | leaves the causal
past of $h$ and of the entire $\gamma$, too. So, the falling observer
will
\emph{not} see the entire future of the Universe. Moreover, he will see
nothing \emph{at all} beyond the shadowed region in Fig.~\ref{fig:Sch}b.
In particular, the last event in $\alpha$'s life observed  by $\gamma$
before the latter submerges into $\mathcal M_*$, is $d$.

\begin{remark}
In  Reissner-Nordstr\"{o}m and Kerr black holes under their event
horizons (which are quite similar to Schwarzschild's) there is another
remarkable surface --- the Cauchy horizon. And that horizon
\emph{does} have the property in discussion: an astronaut falling into the
black hole reaches the Cauchy horizon in a finite proper time and crosses
it in a point $p$ that contains in its causal past the whole ``external
universe". Such an astronaut, indeed, \emph{will be able} to see the
death of stars and galaxies, see, e.~g.,~\cite{Chandrasekhar}.
\end{remark}

\section*{Acknowledgements}
The author was supported in part by RNP Grant No.~2.1.1.6826.


\begin{thebibliography}{A}
\bibitem{MTW}C. W. Misner, K. S. Thorne, and J. A. Wheeler,
\emph{Gravitation} (Freeman, San Francisco, 1973).
\bibitem{Regge}T. Regge, \emph{Cronache dell'Universo}
(Boringhieri, Torino, 1981).
\bibitem{Cher}A. Cherepaschuk, in  \emph{Astronomiya: Vek XXI} ed.~by V. Surdin,
(Fryazino, 2007).
\bibitem{KisLogMes6}A. A. Logunov, M. A. Mestverishvili, and V. V.
Kiselev,  Phys.\ Part.\ Nucl. \textbf{37}, 317 (2006).
\bibitem{LL_II}L. D. Landau and E. M. Lifshitz, \emph{The Classical Theory
of Fields} (But\-terworth-Heinemann, 1980).
\bibitem{konc}A. S. Eddington, \emph{The mathematical Theory of Relativity}
(2-nd Edition. Cambridge, University Press, 1924); \\ V. Fock,
\emph{The Theory of Space, Time, and Gravitation} (Pergamon, New York, 1959).
\bibitem{KS}R. Kantowski and R. Sachs, J. Math. Phys. \textbf{7},
443 (1966).
\bibitem{Chandrasekhar}S. Chandrasekhar,   \emph{The Mathematical Theory
of Black Holes } (Oxford University Press, New York, 1983); \\ B. Carter,
in
\emph{Black Holes - Les Houches 1972}, ed. by C. DeWitt and B.~S.~DeWitt,
(Gordon and Breach Science Publishers, New York, 1973).
\end{thebibliography}
\end{document}